\documentstyle[12pt,twoside]{article}
\def\greaterthansquiggle{\raise.3ex\hbox{$>$\kern-.75em\lower1ex\hbox{$\sim$}}}
\def\lessthansquiggle{\raise.3ex\hbox{$<$\kern-.75em\lower1ex\hbox{$\sim$}}}
\newcommand{\beq}{\begin{equation}}
\newcommand{\eeq}{\end{equation}}
\newcommand{\beqa}{\begin{eqnarray}}
\newcommand{\eeqa}{\end{eqnarray}}
\newcommand{\beqan}{\begin{eqnarray*}}
\newcommand{\eeqan}{\end{eqnarray*}}
\newcommand{\ba}{\begin{array}}
\newcommand{\ea}{\end{array}}

\newcommand{\A}{{\cal A}}

\newcommand{\M}{{\cal M}}

\newcommand{\T}{{\cal T}}

\def\nz{\ifmmode {I\hskip -3pt N} \else {\hbox {$I\hskip -3pt N$}}\fi}
\def\zz{\ifmmode {Z\hskip -4.8pt Z} \else
       {\hbox {$Z\hskip -4.8pt Z$}}\fi}
\def\qz{\ifmmode {Q\hskip -5.0pt\vrule height6.0pt depth 0pt
       \hskip 6pt} \else {\hbox
       {$Q\hskip -5.0pt\vrule height6.0pt depth 0pt\hskip 6pt$}}\fi}
\def\rz{\ifmmode {I\hskip -3pt R} \else {\hbox {$I\hskip -3pt R$}}\fi}
\def\cz{\ifmmode {C\hskip -4.8pt\vrule height5.8pt\hskip 6.3pt} \else
       {\hbox {$C\hskip -4.8pt\vrule height5.8pt\hskip 6.3pt$}}\fi}
\newtheorem{theorem}{Theorem}

\def\au{{\setbox0=\hbox{\lower1.36775ex%
\hbox{''}\kern-.05em}\dp0=.36775ex\hskip0pt\box0}}
\def\ao{{}\kern-.10em\hbox{``}}

\normalsize
\begin{document}
\bibliographystyle{plain}

\begin{titlepage}
\begin{flushright}
\today
\end{flushright}
\vspace*{2.2cm}
\begin{center}
{\Large \bf Time asymptotics for interacting systems }\\[30pt]

Heide Narnhofer  $^\ast $\\ [10pt] {\small\it}
Fakult\"at f\"ur Physik \\ Universit\"at Wien\\

\vfill \vspace{0.4cm}

\begin{abstract}We argue that for Fermi systems with Galilei invariant interaction the time evolution is weakly asymptotically abelian in time invariant states, but not norm asymptotically abelian.Consequences for the existence of time invariant states are discussed.

\\
\hspace{1.9cm}

\end{abstract}
\end{center}

\vfill {\footnotesize}

$^\ast$ {E--mail address: heide.narnhofer@
univie.ac.at}
\end{titlepage}
\section{Introduction}
It is a general belief that interaction is responsible for the approach to equilibrium. However exact control on the asymptotics of time evolution is only given for models that are closely related to free evolution and therefore allow the construction of time invariant states that are far from equilibrium. Therefore it is desirable to search for mathematical properties that hold for interacting systems but not for quasifree systems and in addition offer additional demands for the construction of invariant states.
In \cite{NTS}, \cite{NT} and \cite{BBN} models are offered, for which it could be shown that the only invariant state is the tracial state. In these models the evolution is discrete, therefore without physical interpretation, but we observe, that  on one hand the evolution spreads to infinity, but with additional assumptions on randomness commutators with local operators do not vanish. For systems related to models of physics, spreading to infinity follows if operators converge weakly to their expectation value in the invariant state, i. e. if the time evolution is weakly asymptotically abelian. If they still do not commute with local operators this convergence cannot be improved to norm asymptotic abelianess. Therefore we consider the level of asymptotic abelianess as candidate to indicate different behaviour with respect to thermodynamics.

In this note we consider continuous systems of Fermions interacting by a pair potential that is translation invariant, short ranged and vanishing for particles with large different velocities, but still Galilei invariant. For such a system it was shown in \cite{NT2}, that time evolution is given by a continuous automorphism group on the algebra built by Fermionic creation and annihilation operators. The cutoff for the velocities was necessary for controlling perturbation expansion  in the interaction, but should be of only technical importance for stable interaction where the velocity of the particles  essentially does not increase by the interaction. Due to the relation between time evolution and space translation given by Galilei invariance it is also possible to show in the tracial state or under mild and plausible assumptions on spacial clustering  \cite{N} that in time invariant states with trivial center time correlations inherit the clustering properties from the space correlations and therefore in these states time evolution is weakly asymptotically abelian.
Turning to systems without interaction and the possible extension to systems with quasifree evolution creation operators and annihilation operators anticommute in norm in the limit of large time  provided the one particle hamiltonian has continuous spectrum \cite{N2} and therefore even operator products commute asymptotically in norm. For these systems every quasifree state defined by a one particle density operator commuting with the one particle hamiltonian is time invariant and we have no approach to equilibrium.

In this note we use the fact that the algebra can be considered as crossed product of the even algebra with the automorphism implemented by a unitary constructed from a single creation operator. Using the crossed product formulation has the advantage that it offers an extension to lattice systems \cite{AM} We concentrate on the evolution of this single operator. As odd operator it tends weakly to 0, but a corresponding even operator can be constructed, that for systems without interaction converges in norm, whereas for systems with interaction this arguement fails and norm convergence does not follow. For interacting systems strong convergence is satisfied in the groundstate based on the application of scattering theory in accordance with the results of \cite{J}, but  based on a very plausible analysis  of the time evolution it is violated in the tracial state and only weak asymptotic abelianess remains. We expect that the analysis can be extended to other invariant states.

Finally we discuss why the failing of strong asymptotic abelianess might be of relevance for thermodynamical behaviour.
For odd operators that anticommute it is well know that their expectation value has to vanish in invariant states \cite{R}. Now we increased the class of operators that do not asymptotically commute on the algebraic level. We give hints how these algebraic correlations can give additional restrictions on the invariant states in analogy to the consequences of randomness in \cite{BBN}.

\section{The Fermi-algebra as crossed product}
The Fermi algebra is built by creation and annihilation operators $a^*(f), a(g)$ satisfying
\beq a(f)a(g)+a(g)a(f)=0, \quad a^*(f)a(g)+a(g)a^*(f)=\langle g|f\rangle \eeq
The even algebra $\A_e$ is built by even polynomials in creation and annihilation operators. We pick some $f_0$ which defines the unitary \beq U_0 =a(f_0)+a^*(f_0) \quad U_0^2=1 \eeq
This unitary implements the automorphism $\alpha _0(A_e) =U_0A_eU_0$ on the even algebra and on this algebra is not an inner automorphism and acts freely . The total algebra is now the crossed product of the even algebra with this automorphism, i.e. it consists of pairs of even elements with the multiplication rule
\beq (A_{e1},A_{e2})(B_{e1},B_{e2})=(A_{e1}B_{e1}+ A_{e2}\alpha _0B_{e2},A_{e1}B_{e2}+A_{e2}\alpha _0B_{e1}) \eeq
Especially $U_0$ corresponds in this notation to the operator $(0,1)$ and satisfies $(0,1)(0,1)=(1,0).$ We can construct all odd elements by appropriate multiplication with even elements. Especially the corresponding expression for $a(g)$ reads
\beq a(g)=U_0 (a(f_0 )+a^*(f_0))a(g)\sim (0,1)((a(f)+a^*(f_0))a(g),0)\eeq
We consider the time evolution given by the hamiltonian in dimension $\nu$
\beq
\label {FG}H= \frac{1}{2m}\int dx^{\nu}
\nabla  a^*(x)\nabla a(x)+\int d^{\nu}pd^{\nu}p'd^{\nu}qd^{\nu}q' a^*_{pq}a^*_{p'q'}w(p-p')v(q-q')a_{p'q'} a_{pq}=K+V.\eeq
 Here $a_{pq}=a(W(p,q)f)$ is an annihilation operator smeared with an $f$ that is translated by the one-particle Weyl operators $W(p,q)=e^{i(qP+pX)}.$ As concrete example we take as in \cite{NT} $f$ as a Gauss function so that with appropriate normalization in $\nu $ dimensions  with the notation
$$[a^*(x),a(y)]_+=\delta ^{\nu }(x-y)$$
$$a(f(x))=a(f)=\int d^{\nu} x f(x)a(x)$$
$$a_{pq}=\pi^{-\frac{\nu}{4}}\int d^{\nu }xe^{-\frac{(q-x)^2}{2}+ipx}a(x)=a(|p,q\rangle )$$
where $|p,q\rangle $ are coherent states.

In the following we will omit $\nu $ being of no relevance. This hamiltonian defines an automorphism group that resembles the time evolution of the Fermi system on the lattice in the sense that we can expand in $t$ and can control the convergence \cite{NT}. In the limit $w(p-p')\rightarrow 1$ the interaction aproaches the usual point interaction. The cutoff in the velocities is needed to control the convergence in perturbation expansion. The interaction was adjusted to keep Galilei invariance, which connects space, boost, gauge and time transformations
 \beq\sigma _xa(f(y))=a(f(x+y))\quad \gamma _b a(f(y))=a(e^{iby}f(y))\quad
\nu _{\alpha }a(f)=e^{i\alpha }a(f) \quad
 \tau _t,\eeq
such that
$$ \sigma _x \circ \nu _{\alpha } =\nu _{\alpha } \circ \sigma _x, \quad \gamma _b \circ \nu _{\alpha } =\nu _{\alpha} \circ \gamma _b, \quad \gamma _b \circ \sigma _x=\sigma _x \circ \gamma _b \circ \nu _{-bx}$$
$$ \tau _t \circ \nu _{\alpha }=\nu _{\alpha } \circ \tau _t, \quad \tau _t \circ \sigma _x =\sigma _x \circ \tau _t \quad \tau _t \circ \gamma _b =\gamma _b \circ \tau _t \circ \sigma _{bt} \circ \nu _{-b^2t/2}.$$
 The advantage of the above model is the fact that  the time evolution is related to the space translation by Galilei invariance so that it can inherit weak asymptotical abelianess \cite{NT2},\cite{NT3}, \cite{N}: slightly smearing over the Galilei automorphism adds a contribution of space translations and the decay of the spacial correlation functions is inherited by the time correlation functions. This tells us especially, that no local operator remains strictly local in the course of time.
 The hamiltonian corresponds to even operators, therefore the time automorphism can be considered as an autmorphism on the even algebra that however does not commute with our automorphism $\alpha _0.$ But $ \tau _t  \alpha _0\tau _{-t}  \alpha _0$ is an inner automorphism on the even algebra implemented by an operator $(V_t,0)$. It can be used to extend the time automorphism to the full algebra. Under the assumption that the odd element $(0,1)$ is mapped into the odd element \beq \tau _t(0,1)=(0,V_t)\eeq
 we calculate
 \beq \tau _t (0,1)(A,0)(0,1)=\tau _t(\alpha _0 A,0)= (0,V_t)(\tau _t A,0)(0,V_t)= (V_t \alpha _0 (A)V_t,0)\eeq
 in agreement with the above definition of $V_t.$

 We collect the properties of $V_t$: in a time invariant state that is also space-translation invariant all odd elements converge weakly to $0$ \cite{R}. Therefore
 \beq w\lim _{t\rightarrow \infty }V_t =0\eeq
  Next we observe
  \beq \tau _t(0,1)(0,1)=\tau _t(1,0) =(1,0)= (0,V_t)(0,V_t)=(V_t \alpha _0 V_t,0)\eeq and therefore
  \beq V_t^* - \alpha _0V_t=0. \eeq
  Weak asymptotic abelianess given from Galilei invariance implies that
  \beq  w\lim _{t\rightarrow \infty }(V_t,0)=w\lim _{t\rightarrow \infty }(0,1)(\alpha _0 V_t ,0)(0,1)=w\lim _{t\rightarrow \infty }(\alpha _0 V_t ,0)=w\lim _{t\rightarrow \infty }(V_t^*,0)\eeq

\section{Asymptotics of the time evolution}
For free systems respectively quasifree systems
\beq \tau _t a(f)=a(e^{iht}f)\eeq
Using the anticommutation relation
\beq \lim _{t\rightarrow \infty }a^*(e^{iht}f)a(g)+a(g)a^*(e^{iht}f)= \lim _{t\rightarrow \infty }\langle g|e^{iht}f\rangle =0\eeq
provided the one-particle hamiltonian $h $ has absolutely continuous spectrum. This implies that
\beq  \lim _{t\rightarrow \infty} [ \tau _tA_{e1},A_{e2}]=0 \quad  \lim _{t\rightarrow \infty}[ \tau _tA_e,A_o] \eeq

With respect to the crossed product construction we can reproduce the result by arguing
\beq w\lim _{t\rightarrow \infty }V_t =\lim _{t\rightarrow \infty }(a(e^{iht}f_0)+a^*(e^{iht}f_0))(a(f_0)+a^*(f_0)=0\eeq
But in addition for the quasifree evolution straight forward calculation gives
 \beq V_t-V_t^*=c_t\eeq
 where $c_t$ is just a number. This number  tends weakly to $0$, but as a number therefore also in norm in agreement with (14).
 This argument only works for quasifree evolution and we have to look for arguments how interaction changes the convergence properties.
 \begin{theorem}
 Assume $\omega $ is a time invariant state for which the GNS representation is cyclic and separating and in which \beq
  st-lim _{t\rightarrow \infty }[a(f) \tau _ta(g)^*\
 \pm \tau _ta(g)^*a(f)]=0 \eeq
 Then $$ st-lim _{t\rightarrow \infty }V_t V_t=\pm 1$$
 Let in the GNS representation $$ U|\Omega \rangle =(a(g)+a(g)^*)|\Omega \rangle = U'|\Omega \rangle $$
 with $U'$ the corresponding operator in the commutant and $e^{iHt}$ implementing the time automorphisms satisfying $$e^{iHt}|\Omega \rangle $$ then
 \beq st-lim _{t\rightarrow \infty }e^{iHt}UU'e^{-iHt}=\pm 1. \eeq
 \end{theorem}
Proof: Since
\beq st-\lim _{t\rightarrow \infty}V_t V_t=\pm 1\eeq  the eigenvalues of $V_t$ tend to $\{1,-1 \}$ respectively to $\{i,-i \}$ and therefore the eigenvalues of $V_tV_t$ to $1$ or $-1$. With $A'$  operators in the commutant and $A'|\Omega \rangle $ dense in the Hilbertspace
\beq st-lim _{t\rightarrow \infty }V_t V_tA'|\Omega \rangle =A'Ue^{iHt}U^{-iHt}Ue^{iHt}Ue^{-iHt}|\Omega \rangle =A'Ue^{iHt}Ue^{-iHt}Ue^{iHt}U'|\Omega \rangle \eeq
$$=A'Ue^{iHt}UU'e^{-iHt}Ue^{iHt}|\Omega \rangle =A'Ue^{iHt}UU'e^{-iHt}U|\Omega \rangle $$
where we have used the effect of $e^{iHt}$ and of $U$ on $|\Omega \rangle $ and thus turned the product of operators in the algebra into the time evolution of one operator, though now belonging to the bounded operator over the Hilbertspace and not any more to the initial algebra.
In order to show that creation and annihilation operators do not anticommute asymptotically in norm it suffices to show that theorem 1 is not satisfied for just one invariant state. We will do this in the tracial state.

\section{The tracial state as the Vacuum state over an extended algebra}

In \cite{J} it was shown that applying the results of scattering theory in the vacuum state the time automorphism also for interacting systems is if not in norm so strongly asymptotically abelian. Scattering theory does not exist in temperature states \cite{NRT}, and in general we do not have a replacement to control long time behaviour. This is different for the tracial state, that is invariant under all automorphisms and allows a concrete construction of the GNS-representation that will enable us to perform estimates also for interacting systems.
\begin{theorem} With the observable algebra $\A(a)$ built by creation and annihilation operators $a(f),a^*(f) $ the GNS representation of the tracial state coincides with the vacuum state over an extended Fermi-algebra built by creation and annihilation operator $A(f),A^*(f),B(f), B^*(f)$ connected with the initial algebra by a Bogoliubov transformation
\beq a(f)=\frac{1}{\sqrt2}(A(f)+ B^*(\bar{f})), \eeq
With
\beq b(f)=\frac{1}{\sqrt2}(A(f) -B^*(\bar{f}))\eeq
\beq [a(f),b(g)]_+=0,[a(f), b^*(g)]_+=0,[b(f),b(g)]_+=0,[b(f)^*,b(g)]_+=\langle g|f\rangle \eeq
The even algebra $\A(b)_e$ commutes with $\A(a).$
Let $|\Omega \rangle $ be the vacuum vector for $\A(A,B)$ and $H_e$ respectively $H_o$ be the subspaces built from $|\Omega \rangle $ by even respectively odd polynomials in $A,B,A^*,B^*$ and let \beq W |\Psi _e\rangle =|\Psi _e \rangle \quad W|\Psi _o \rangle=-|\Psi _o\rangle, \quad |\Psi _{e,o}\rangle \in H_{e,o} \eeq
then $Wb(f), b^*(g)W$ commute with $\A (a)$ and create the commutant $\A(a)'$ \end{theorem}
Proof: Both the vacuum state over $\A(A,B)$ and the tracial state over $\A(a)$ are determined by the two point function \beq 2\omega (a(f)a^*(g))=\langle g|f\rangle =\frac{1}{2}\omega (A(f)+B^*(\bar{f})(A^*(g)+B(g)))\eeq
Further with $J$ the modular conjugation
\beq a(f)|\Omega \rangle =Ja(\bar{f})|\Omega \rangle =\frac{1}{\sqrt2} B^*(\bar{f})|\Omega \rangle =Wb(f)|\Omega \rangle, \quad Ja(\bar{f})J=Wb(f). \eeq
The time evolution is implemented by an operator satisfying $H=-JHJ, H|\Omega \rangle =0$ . Acting on $\A (a)$ it is implemented by a sequence of local operators $h_{\Lambda }$ that we can express term by term by $A(f),B(f),A^*(g)B^*(g)$ First we concentrate on quadratic terms:
\beq 2a^*(f)a(g)=A^*(f)A(g)+B(\bar{f})A(g)+A^*(f)B^*(\bar{g})+B(\bar{f})B^*(\bar{g})\eeq
These terms do not annihilate the vacuum. However the effect on $\A(a)$ remains unchanged if we add counterparts from the commutant, including c-numbers:
\beq b^*(f)b(g)=A^*(f)A(g)-B(\bar{f})A(g)-A^*(f)B^*(\bar{g})+B(\bar{f})B^*(\bar{g})\eeq
Therefore we choose as contributions to the hamiltonian
\beq a^*(f)a(g)+b^*(f)b(g)-\langle g|f\rangle =A^*(f)A(g)-B^*(\bar{g})B(\bar{g})\eeq
Notice that the necessary c-number renormalization tends to infinity in agreement with the fact that the time evolution is not an inner automorphism of $\A(a).$
We observe that a quasifree time evolution on $\A(a)$ becomes a quasifree time evolution also on $\A(A,B)$ and decouples between $\A(A)$ and $\A(B).$
We turn to the contributions of the interaction. They are a polynomial of forth order in creation and annihilation operators. Expressed by $\A(A,B)$ they contain contributions with four creation operators, with three creation operators with two creation operators and with one creation operator. The contribution from the commutant necessarily has to annihilate the term with four creation operators in order to satisfy $H|\Omega \rangle =0.$
This holds if we choose \beq a^*(f)a^*(g)a(g)a(f)-b^*(f)b ^*(g)b(g)b(f)\eeq and observe that it contains terms like $A^*(f)B^*(f)B^*(g)B(g)$
where we use that for the expression following from the contributions of $b$ the terms with four, two and zero creation operators cancel and only the terms with three and one remain, where in addition the annihliation operator has to be commuted to the right which gives an additional quadratic contribution.
We observe two facts: On one hand the time evolution on the algebra $\A(b)$ is given by the kinetic part together with an interaction similar as for the observable algebra but with opposite sign (Compare (30) and (31)). If we believe that the sign of the interaction is relevant for stability properties of the system, we have to expect that the resulting instability in the commutant can have consequences also in the algebra. That this is true can be seen by turning to the time evolution for the extended algebra in its vacuum state. The time evolution is not any longer number preserving and we have to evaluate its consequences also for the observable algebra.

According to theorem (1) we have to evaluate the convergence of (20). We express it in terms of $A,B,$ suppressing the dependence on $f$
\beq UU' = \frac{1}{2}(A+B +A^*+B^*)(A+B-A^*-B^*)=\frac{1}{2}(A^*+B^*)(A+B)-\frac{1}{2}(A+B)(A^*+B^*)\eeq
For quasifree gauge invariant evolution this operator converges strongly to $-1$ in the vacuum representation provided the hamiltonian in the one-particle space has continuous spectrum. For interacting systems we have to show that
\beq \lim _{t\rightarrow \infty}\frac{1}{4}\langle \omega|(A+B)e^{iHt}((A^*+B^*)(A+B)-(A+B)(A^*+B^*))e^{-iHt}(A^*+B^*)|\Omega \rangle \neq -1\eeq
We analyse the effect of the hamiltonian: The hamiltonian breaks gauge symmetry. As a consequence the potential acting on a one particle state the one particle contribution remains unchanged whereas a pair of additional  particles is created with the weight given by the potential provided the corresponding place is empty. Repeating the action of the potential the additional pair is annihilated so that we return to the initial state.  Therefore without quasifree part in the hamiltonian the action is periodic and the excitation stays localized. Adding the quasifree part that is responsible for weak asymptotic abelianess  it separates the additional pair locally, especially since the quasifree part acts for the two kind of particles in different direction. Therefore new single excitations without neighbors are created and enable the creation of additional pairs. In addition the fact that the quasifree part acts in different direction guarantees that some particle excitations remain locally. Therefore in the average the particle number will increase linearly in time. Also the delocalisation increases linearly in time so that we obtain in the mean a positive particle density. But this implies that $(A^*_f +B^*_f )(A_f +B_f )$ as part of the number operator cannot totally annihilate this many-particle state.

 Strong asymptotic anticommutativity would demand that
\beq \langle \omega |(A+B)e^{it(H_0+V)}(A^*+B^*)(A+B)e^{-it(H_0 +V)}(A^*+B^*)|\Omega \rangle =0, \eeq $$ \lim _t||(A+B)e^{-it(H_0 +V)}(A^*+B^*)|\Omega \rangle||=0$$

which according to our consideration will not be satisfied. Therefore

\begin{theorem}For interacting Galilei invariant systems the time evolution in the tracial state that violates (34) is weakly asymptotically abelian but not strongly asymptotically abelian. \end{theorem}

\section{Consequences}
Both strong and weak asymptotic abelianess  imply that local operators do not remain local for increasing time. Comparing with a strictly local evolution where we can construct invariant states by choosing locally an invariant state depending on the position and take arbitrary products of these states, nonlocality produces constraints between far separated regions. We observe, that there are more constraints if the time evolution is only weakly asymptotic abelian:   Choosing a subalgebra $\M_2$ with dimension $2$ of our observable algebra we can write \beq \A(a)=\M_2 \otimes \A _M\eeq
With $m\in \M_2$ we consider $(\tau _t m)_{ik}=$ as element in $\A_M$. If the time automorphism is norm asymptotially abelian then $\lim _t [(\tau _t m)_{11}-(\tau _t m)_{22}]=0,\lim _t (\tau _t m)_{12}=\lim _t (\tau _t m)_{21}=0.$
If they do not commute then $||(\tau _t m)_{12}|| > 0$ for arbitrary large $t$.  An invariant state is given by functionals $ \omega _{ik} $ over $\A_M$ with $\omega (m_{12})=\omega _{12}(1).$
$ (\tau _t m)_{12}$ varies with $t$ and therefore implies constraints for $ \omega _{ik} $ If this variation is random which is the case for the annihilation operator $m=a(f)$ and for all operators in the case of \cite {BBN} then it follows that $\omega _{12}=0=\omega (m_{12})$ and we obtain a condition on the local level. In general we expect that the correlations between $\omega _{ik}$ that are present in the case of weak asymptotic abelianess will give bounds on $\omega _{12} (1).$

That these bounds are more restrictive for weak asymptotic abelianess than for norm asymptotic abelianess seems plausible. But in addition we can observe that they are more demanding for dynamical stability \cite{HKTP}.We study the reaction of an invariant state to a local perturbation of the dynamics. Dynamical stability demands that we can find a state invariant under these new dynamics in the folium, i. e. the perturbation is unable to create in the course of time infinite energy change.

Galilei invariance allowed that spacial clustering is inherited by time clustering. However the proof fails for multiclustering, and in fact for norm asymptotically abelian time evolution \beq \lim _{t\rightarrow \infty }\omega (AB_tCD_t) = \omega (AC)\omega (BD)\omega (AB_tCD_t) = \omega (AC)\omega (BD) \eeq
whereas for a time evolution that is weakly asymptotically abelian \beq \lim _{t\rightarrow \infty }|\omega (AB_tA^*B^*_t)| < \omega (AA^*)\omega (BB^*). \eeq
Let us assume that  states extremely invariant under space translation cluster uniformly and this clustering is inherited by time translation
\beq \lim _{t\rightarrow \infty } P_{\Lambda_2 ^C}e^{iHt}P_{\Lambda _1}=|\Omega \rangle \langle \Omega | \eeq
where $\Lambda_2$ containing $\Lambda_1$ are local regions,$\Lambda^C$ the complement of $\Lambda$ and $P_{\Lambda }$ the projection operator onto $\A_{\Lambda }|\Omega \rangle.$
Under this assumption together with norm asymptotic abelianess
\beq \lim_{t_j -t_i \rightarrow \infty } \omega (\tau _{t_1} A_1\tau _{t_2} A_2...\tau _{t_k} A_k)=\omega (A_j)\omega_{i\neq j} (\Pi \tau _{t_i} A_i)\eeq
independently of the ordering and the size of $t_i$ and we can control multiclustering. This makes it possible to control local perturbations
\beq \langle \Omega |Ae^{it(H+\lambda V)}Be^{-it(H+\lambda V)}A^*|\omega \rangle .\eeq By expanding in the coupling constant and taking the limit $t\rightarrow \infty $ we can obtain an invariant state. In KMS-states where the time automorphism coincides with the modular automorphism we can construct this invariant state by the time ordered expression
\beq |\Omega _V\rangle = \T \int _0^{\beta/2} d\gamma e^{\tau_{i\gamma }V}|\Omega  \rangle \eeq
where now the integration runs over a finite region and does not refer to asymptotic behaviour, therefore can also be applied for systems that are just weakly asymptotically abelian.
An other possiblity to construct the invariant state is looking for the vector in the natural cone satisfying
\beq (H+\lambda (V-JVJ))|\Omega _V\rangle =0=|\Omega +\sum \lambda ^n B_n\rangle \eeq which is solved by using the modular operator $M$ and expanding in $\lambda $
\beq B_1|\Omega \rangle =-\frac{1}{H}(V-JVJ)|\Omega \rangle =-\frac{1}{H}(1-e^{-M/2}) V|\Omega \rangle \eeq
\beq B_n|\Omega \rangle =\frac{1}{H}B_{n-1}|\Omega \rangle \eeq
Considering
\beq \lim _{\epsilon \rightarrow 0} \frac{-i}{H+i\epsilon} =\lim _{\epsilon \rightarrow 0}\int_0^{\infty } dt e^{it(H+i\epsilon)} \eeq
we are dealing with an unbounded operator $\frac{1}{H}$, but from clustering in time it follows that $(V-e^{-M/2}V)|\Omega \rangle $ belongs to its domain. However for $B_n|\Omega \rangle$ we would need higher order of time correlation functions that for weakly asymptotically abelian systems are not available.

Therefore we have to be aware that differently as for KMS-states local perturbations of the dynamics can lead to states that are not normal with respect to the initial state, i.e. that already the local perturbations can create infinite energy in the course of time. We observe different behaviour with respect to the degree of abelianess, but also the special role of KMS-states. Extending the local perturbation to global perturbation as it is possible for KMS-states without violating the KMS-property with respect to the new dynamics is even more problematic. All these are indications of the sensibility of weakly asymptotically systems and increasing restrictions for invariant states.

\section{Conclusion}
We have used the crossed product construction offered by the difference of even and odd operators for a Fermionic system to show that the passage of asymptotically weak abelianess of odd operators to asymptotic abialness for even operators asks for a detailed balance between weakly converging operators. This balance is guaranteed for quasifree evolution. In the tracial state ( and probably with appropriate variations in all other states that are cyclic and separating ) the time evolution considered as the limit of local time evolutions differs for the algebra and its commutant in the time direction, if the evolution is quasifree, but in addition in the sign of the interaction for interacting systems. We argue that we keep weak asymptotic abelianess due to Galilei invariance of the time evolution but that strong asymptotic abelianess is destroyed due to the violation of number conservation for the rotated particles. A detailed analysis of an evolution in Fock space without particle number conservation would be desirable. Finally we search for arguments why weak asymptotic abelianess might reduce the possibility for time invariant states as suggested by examples of discrete time evolution. On one hand it offers additional areas of randomness that create restrictions on space correlations. On the other hand we refer to the demand of dynamical stability i.e. the existence of invariant states under local perturbations without leaving the follium. This holds for KMS-states, but otherwise the necessary analycity properties get out of control for interacting systems. This might be compared with the fact, that for interacting systems the time evolution is only analytic with finite range, we have different analyticity behaviour on the algebraic level for system with interaction and without interaction.

We are optimistic that further analysis that concentrates on long range effects in time on the algebraic level and their effect on long range correlations in space for invariant states
can lead to deeper understanding of the importance and the power of interaction with respect to thermodynamics.

\bibliographystyle{plain}

\end{document}